\documentclass[epjST]{svjour}
\usepackage{graphics}
\usepackage{epsfig}
\usepackage{graphicx}
\usepackage{epsfig,color}
\usepackage{cite}
\usepackage[english]{babel}
\usepackage{amsmath}
\usepackage{floatrow}
\usepackage[label font=bf]{subfig}
\usepackage{caption}
\usepackage{siunitx}
\usepackage{xcolor}

\begin{document}
\title{Rhythmogenesis in the mean field model of the neuron-glial network}
\author{Nikita Barabash\inst{1,2}\fnmsep \and Tatiana Levanova\inst{2}\fnmsep\thanks{\email{tatiana.levanova@itmm.unn.ru}} \and Sergey Stasenko\inst{2}\fnmsep\thanks{\email{stasenko@neuro.nnov.ru}} }
\institute{Volga State University of Water Transport, Nizhny Novgorod, 603950, Russia \and Lobachevsky State University of Nizhny Novgorod, 23 Prospekt Gagarina, Nizhny Novgorod, 603950, Russia}
\abstract{
Despite the fact that the phenomenon of bursting activity is important for functioning of living neural networks, the mechanisms of its origin are still not clear. In this paper, we propose a new phenomenological model that can explain the mechanisms of the formation of bursting activity based on short-term synaptic plasticity, recurrent connections, and neuron-glial interactions. We show that neuron-glial interactions can induce bursting activity. The bifurcation scenarios of emergence of bursting activity are in the focus of the paper. Proposed study is important for understanding of the complex dynamics in neural networks.
} 
\maketitle
\section{Introduction}
\label{intro}

The complex collective dynamics of a neural networks include various patterns of activity, the most interesting of them being bursting activity. A burst consisting of a short, high-frequency spike sequence is more likely to pass through a synapse than a single spike, and the likelihood of a postsynaptic spike is increasing correspondingly. Both the number and temporal structure of spikes in a burst provide an encoding space and basis for temporal integration of individual neurons \cite{Segev2004,Hulata2004}.

Bursting activity associated with the collective dynamics of neurons plays an important role in the functioning of the neural system. It underlies both the normal physiology of the brain (e.g. with the rhythmogenesis) and brain pathology, in particular epilepsy. Also, in a number of previous studies it was shown that bursting activity makes a significant contribution to the development of the visual system \cite{Meister1991}, sensory processing \cite{Krahe2004}, neural transmission \cite{Salinas2001}, learning and memory \cite{Axmacher2006}.

Many attempts have been made to study this phenomenon because of its exceptional importance. First of all, it is necessary to note a number of experimental studies of bursting activity in cultures of neurons grown on multielectrode arrays \cite{Wagenaar2006}. In those studies researchers changed culture properties, such as density \cite{Ito2010, Ivenshitz2010}, size \cite{Wilson2007} or the stage of development \cite{Biffi2013, VanPelt2004}, as well as pharmacological conditions \cite{Penn2016}. Also, different dynamical properties of bursting activity, such as self-adjusting complexity \cite{Hulata2004}
, have also been studied in detail.
The interest of researchers in this topic is due not only to the possibilities of encoding information, but also to the potential possibility of creating a living neural chip \cite{Baruchi2007} with predetermined functions. However, this goal is far from being fully achieved.

Results obtained in described biological experiments can be explained using mathematical modeling. 
In order to understand mechanisms of bursting activity, a number of mathematical models \cite{Masquelier2013, Maheswaranathan2012} have been proposed recently. In particular, in \cite{Markram1996,Blitz2004} it was shown that short-term synaptic plasticity (STSP) can be viewed as a possible synaptic mechanism for the formation of bursting activity. 

STSP is a temporary (on a time scale from a few seconds to minutes) change (increase or decrease) in the strength of a synaptic connection in response to a short-term stimulus. This fact was first discovered in the studies of synaptic transmission in neocortex \cite{Markram1996,Tsodyks1997,Thomson1994}. The effect of STSP is based on the accumulation of $Ca^{2+}$ in presynaptic terminals in response to short-term exposure, which, in turn, leads to a change in the probability of neurotransmitter release due to modulation of exocytosis \cite{Citri2008, Zucker2002}. A short-term change in synaptic strength leads to a change in the activation of postsynaptic neurons which in its turn results in modulating network dynamics and cognitive processes 
\cite{Wang2006}.

It is worth noting that there is an increasing evidence that glial cells (astrocytes) may also play a role in the regulation of synaptic dynamics \cite{Sibille2015, Haydon2015, Halassa2010, Jolivet2015}.  However, most of the previous studies of the formation of bursting activity have not taken into account the influence of glial cells.

The fact that astrocytes influence synaptic transmission led to the hypothesis of the so-called tripartite synapse. This term was first introduced in \cite{Araque1998} on the basis of biological evidence for the existence of a bidirectional interaction between neurons and astrocyte. According to this hypothesis, a neurotransmitter, a neuroactive substance released during synaptic signal transmission, can reach metabotropic glutamate receptors on the membrane surface of glial cells, which lead to their activation and subsequent release of gliotransmitters. In its turn,  gliotransmitter (specifically, glutamate) can change the probability of neurotransmitter release by the presynaptic neuron, thereby forming a feedback loop \cite{Araque1999}.

In this paper, we propose a new mathematical model to describe the formation of bursting activity. Our model takes into account the main features of neuron-glial interactions. The proposed model is based on the Tsodyks-Markram \cite{Mongillo2008} model.  The main attention in the present study is paid to the mechanisms of the formation of bursting activity. The results of our modeling show that neuron-glial interactions can produce bursting activity. The presented research will help to contribute to the understanding of the complex dynamics of neural networks.

\section{The model} \label{sec:model}

Tsodyks and Markram propose a simplified phenomenological model of STSP in \cite{tsodyks1998neural}. Original Tsodyks-Markram model describes the deterministic behavior of a population of identical excitatory neurons using a three-dimensional system of ODE with state variables $E(t)$, $x(t)$ and $u(t)$ as follows. 

\begin{equation}
\begin{array}{l}
\tau\dot{E}= -E+\alpha\ln\left[1+\exp\left(\dfrac{JuxE+I_0}{\alpha}\right)\right],\\
\\
\dot{x}=\dfrac{1-x}{\tau_D}-uxE,\\
\\
\dot{u}=\dfrac{U-u}{\tau_F}+U(1-u)E,\\

\end{array}
\label{sys:TMmodel}
\end{equation}

Here $E(t)$ is the average neuronal activity of the excitatory population (in Hz) at any given time. Parameter $I_{0}$ is the inhibitory input received from the recurrent network $E(t)$. The $\alpha$ parameter determines the threshold for increasing the average neuronal activity of the excitatory population.
It should be noted that the positive feedback $JuxE$ contains both structural $J$ and synaptic factors $ux$.

Variable $x(t)$ models the amount of available neurotransmitter (glutamate). According to this model, presynaptic resources are finite and each of them can be either available or non-available to be released. The overall fraction of available neurotransmitter is $x(t)$, and the fraction of non-available neurotransmitter is $(1-x(t))$. In the case of network activity $E(t)>0$ neurotransmitter is consumed, which results in short-term synaptic depression. Term $u(t)E(t)$ corresponds to the consumption rate. The parameter $\tau_D$ represents the spontaneous recovery time from the depressed state.

The variable $u(t)$ describes the change in the probability of release of the neurotransmitter from the presynaptic terminal. A fraction $1-u(t)$ has a low probability of being released; a fraction $u(t)$ has a high-probability of being released, respectively. Also, the term $UE(t)$ is the transition rate from low-probability releasable state to high-probability releasable state. In the original model parameter $U$ is a constant that encodes the baseline level of $u(t)$. Time constant $\tau_F$ describes the characteristic time of the facilitation.

\begin{figure*}[t]
    \centering
    \includegraphics[width=0.7\textwidth]{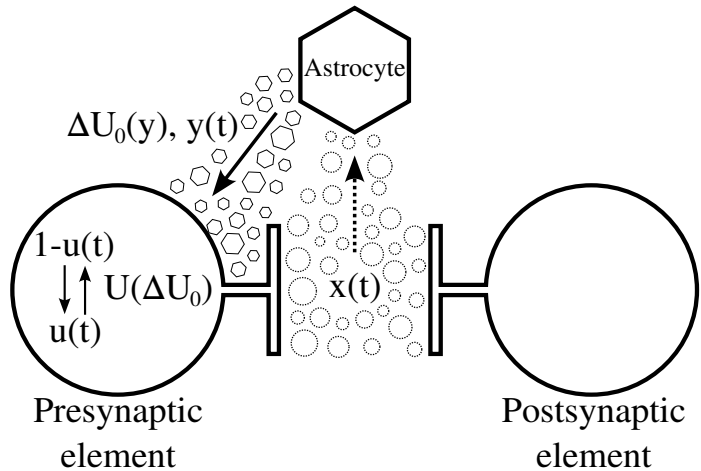}
    \caption{Overall scheme representing the physical phenomenon: astrocyte modulates synaptic transmission at tripartite synapse. During synaptic activity the presynaptic element releases a neurotransmitter (glutamate) $x(t)$, which activates receptors at the postsynaptic element and influences an astrocyte. Activation of the corresponding 
    receptors on the astrocyte membrane leads to calcium elevations. This process helps to control neuronal excitability and synaptic transmission through the calcium-dependent release of gliotransmitter $y(t)$ that alters the probability of release of neurotransmitters at the presynaptic terminal. Here $\Delta U_0(y)$ is an influence of the astrocyte on neurotransmitter release probability $U(\Delta U_0(y))$, $1-u(t)$ -- low-probability releasable state, $u(t)$ -- high-probability releasable state. }
\label{fig:scheme}
\end{figure*}

To describe the dynamics of the glial cell (astrocyte), we used the approach proposed by the authors in \cite{Lazarevich2017}. 
The schematic representation of the biological mechanism of neuron-glial signaling via tripartite synapse is presented in Fig.~\ref{fig:scheme}. Mathematically these interactions were described as follows. The system \eqref{sys:TMmodel} was supplemented with an additional equation for the variable $y(t)$, which describes the change in the concentration of the gliotransmitter released as a result of a cascade of biochemical reactions during the neuron-glial interaction. Also, we took into account the dependence of the baseline level of $u(t)$, $U$, on the variable $y(t)$. The resulting four-dimensional system of ordinary differential equations can be written in the following form:
\begin{equation}
\begin{array}{l}
\tau\dot{E}=
-E+\alpha\ln\left[
1+\exp\left(\dfrac{JuxE+I_0}{\alpha}\right) 
\right],\\
\\
\dot{x}=\dfrac{1-x}{\tau_D}-uxE,\\
\\
\dot{u}=\dfrac{U(y)-u}{\tau_F}+U(y)(1-u)E,\\
\\
\dot{y}=-\dfrac{y}{\tau_y}+\beta \sigma(x),\\ 
\end{array}
\label{sys:model}
\end{equation} 

The characteristic relaxation time of the gliotransmitter $\tau_{y}$ is 1 s, $\sigma(x)$ is a sigmoid function of the form:
\begin{equation}
\sigma(x) = \frac{1}{1 + e^{-20(x-x_{thr})}},
\label{sys:sigmaY}
\end{equation} 
where $x_{thr}$ is the astrocyte activation threshold. The release of the gliotransmitter results in a change in the baseline probability of neurotransmitter release. Experimental studies show that, depending on the type of presynaptic receptors, the probability of release in the presence of a gliotransmitter can either increase (potentiation) or decrease (depression). In our model, the change in release probability in the presence of a gliotransmitter is described as follows:
\begin{equation}
   U(y) = U_0  + \frac{\Delta U_0}{1 + e^{-50(y-y_{thr})}},
\label{sys:UY}
\end{equation}
where $U_0$ is the probability of neurotransmitter (glutamate) release in the absence of astrocytic influence, $\Delta U_0$ is the change in the release probability due to the action of the gliotransmitter on the presynaptic terminal, and $y_{thr}$ is the threshold value that determines the change in the release probability due to effects of gliotransmitter on the presynaptic terminal.

Parameters of the original Tsodyks-Markram model, as well as corresponding parameters of the proposed model in the present study can be identified with biophysical variables, that can be found experimentally. 

In this paper, the parameter $I_0$ was chosen as the control one. The remaining parameters were fixed and took the following values. We used parameters of neural activity that are essentially typical parameters of the Tsodyks-Markram model: $\tau=0.013$, $\tau_{D}=0.15$, $\alpha= 1.5$, $\tau_{F} = 1$, $J = 3.07$. The neurotransmitter and gliatransmitter parameters were chosen in accordance with the tripartite synapse model proposed earlier in 2012 \cite{Gordleeva2012} and 2017 \cite{Lazarevich2017}: $U_0= 0.23$, $\Delta U_0 =0.305$, $\tau_{y} = 1.8$, $\beta = 0.4375$, $x_{thr}=0.9$, $y_{thr}=0.5$.

\section{Bifurcations and change in temporal patterns of population activity}
\label{sec:bif}

Now let us study mechanisms of rhythmogenesis in the proposed model using methods of nonlinear analysis and bifurcation theory.

\begin{figure*}[t]
    \centering
    \includegraphics[width=0.32\textwidth]{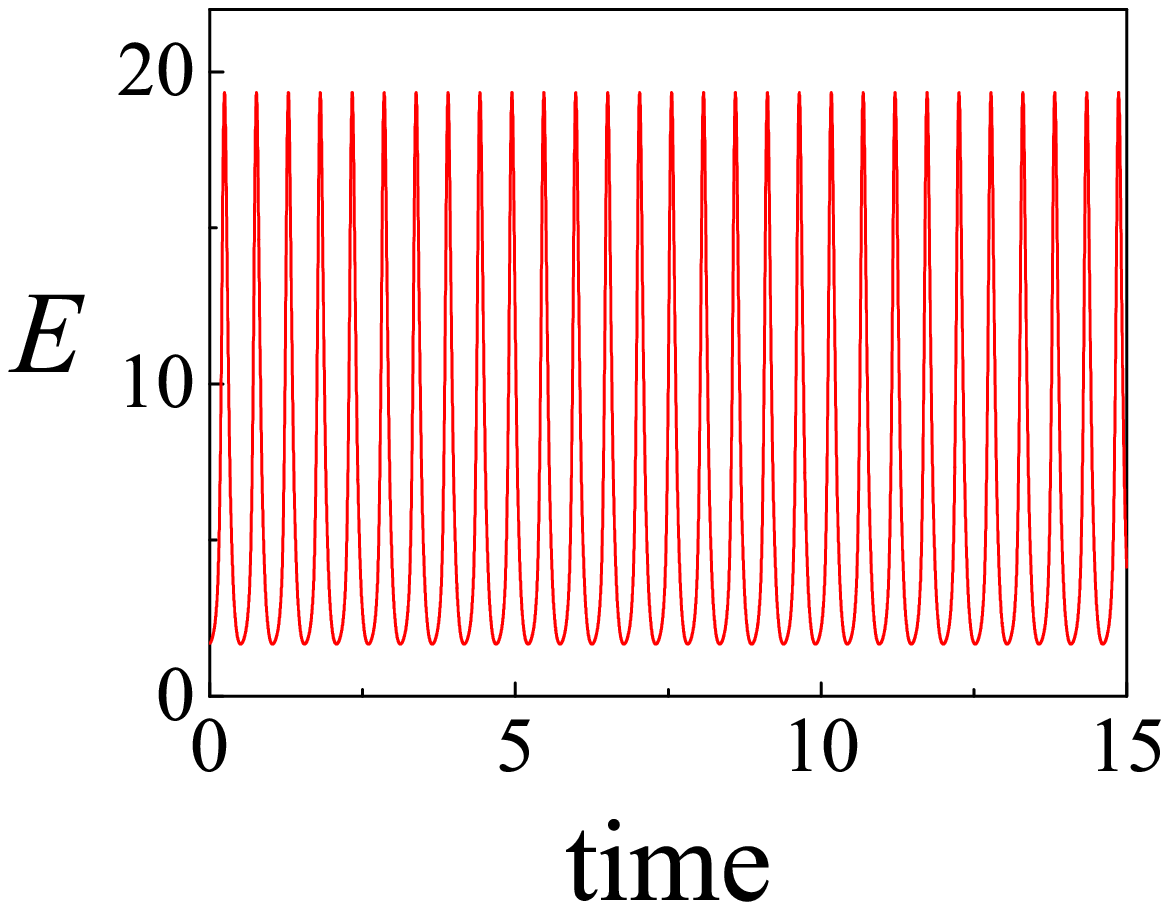}
    \includegraphics[width=0.32\textwidth]{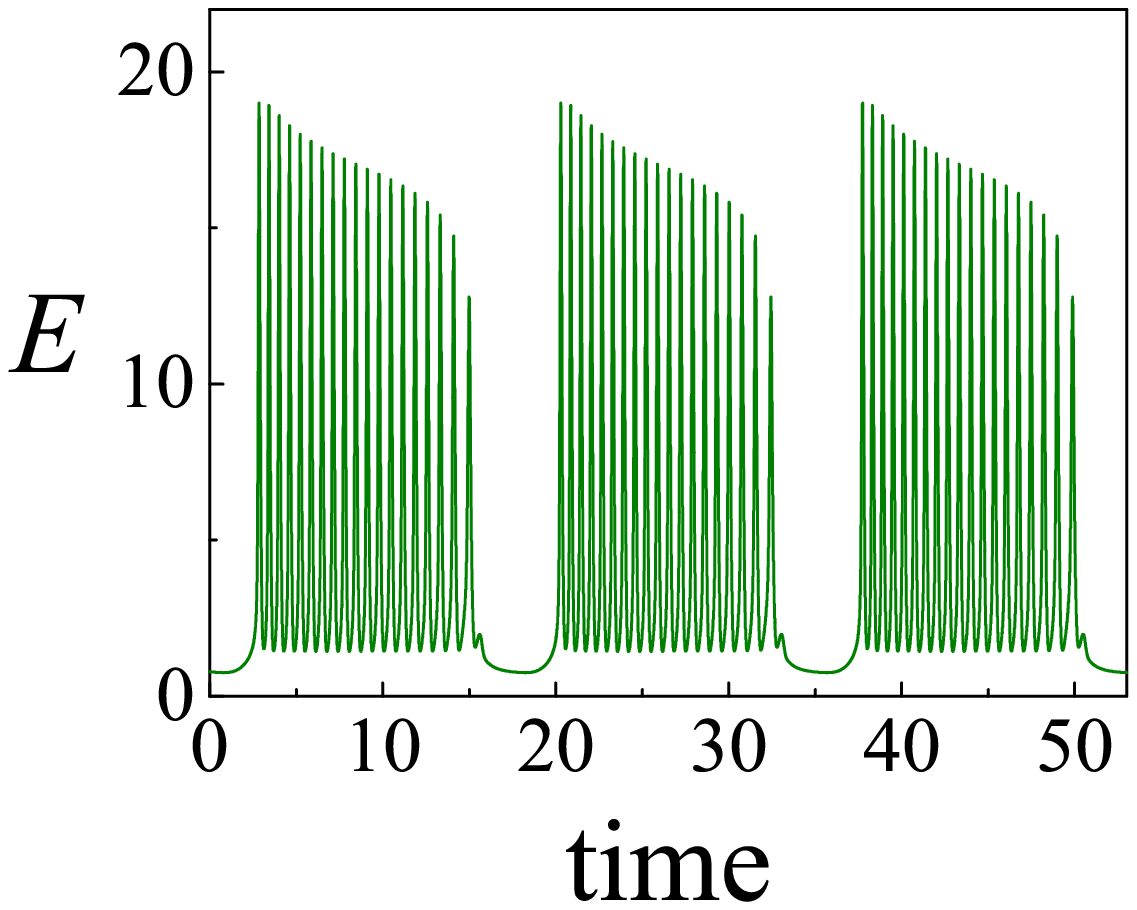}
    \includegraphics[width=0.32\textwidth]{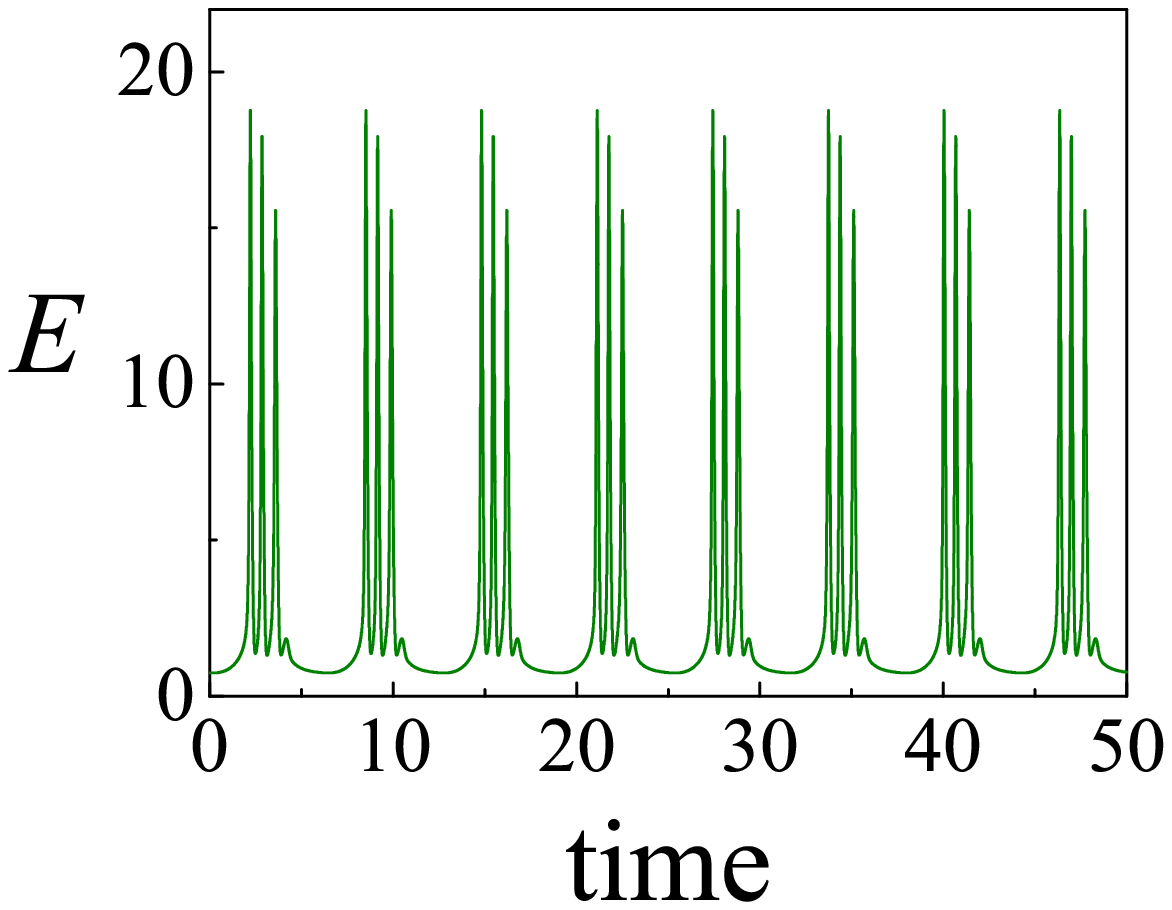}
    \caption{ Different temporal patterns of population activity of the system~\eqref{sys:model}. (left panel) Regular oscillatory (spiking) activity at $I_{0}=-1.42$, which corresponds to a one-loop stable limit cycle (red trajectory in Fig.~\ref{fig:bif_diag}(b)). (middle panel) Bursting activity at $I_{0}=-1.45$. (right panel) Bursting activity at $I_{0}=-1.48$, which corresponds to a multi-loop stable limit cycle (green trajectory in Fig.~\ref{fig:bif_diag}(c)). }
\label{fig:realizations}
\end{figure*}

In dynamical terms, the developed model demonstrates a rich set of patterns of population activity -- from trivial ones (stable equilibrium and one-loop limit cycle that correspond to the excitatory state and tonic spiking), to multi-loop periodic orbits and irregular complex motions corresponding to regular and irregular bursting dynamics (see Fig. \ref{fig:realizations}).

The change of temporal patterns of population activity in the system~\eqref{sys:model} is determined by the bifurcations of its equilibrium states and limit cycles.

Unfortunately, analytical study of these bifurcations presents great difficulties since the right hand sides of the system~\eqref{sys:model} include essentially non-linear functions such as sigmoids $\sigma(x)$, $U(y)$ and logarithm $\ln[1+\exp(\cdot)]$.
This leads to transcendental equations for equilibrium states 
\begin{equation*}
\begin{array}{rl}
E=\alpha\ln\left[
1+\exp\left(\dfrac{JuxE+I_0}{\alpha}\right) 
\right],&\quad
x=\dfrac{1}{1+\tau_DuE},\\
\\
u=U(y)\dfrac{1+\tau_FE}{1+\tau_FEU(y)},&\quad
y=\tau_y\beta \sigma(x),
\end{array}
\end{equation*}
which in general do not allow analytical solution.
This circumstance makes it difficult to carry out a standard local bifurcation analysis of equilibrium states. Therefore, we conducted a numerical study.
The results discussed below were obtained by us using a package for numerical bifurcation analysis MATCONT \cite{Matcont}.

\begin{figure*}
    \centering
    \includegraphics[width=0.55\textwidth]{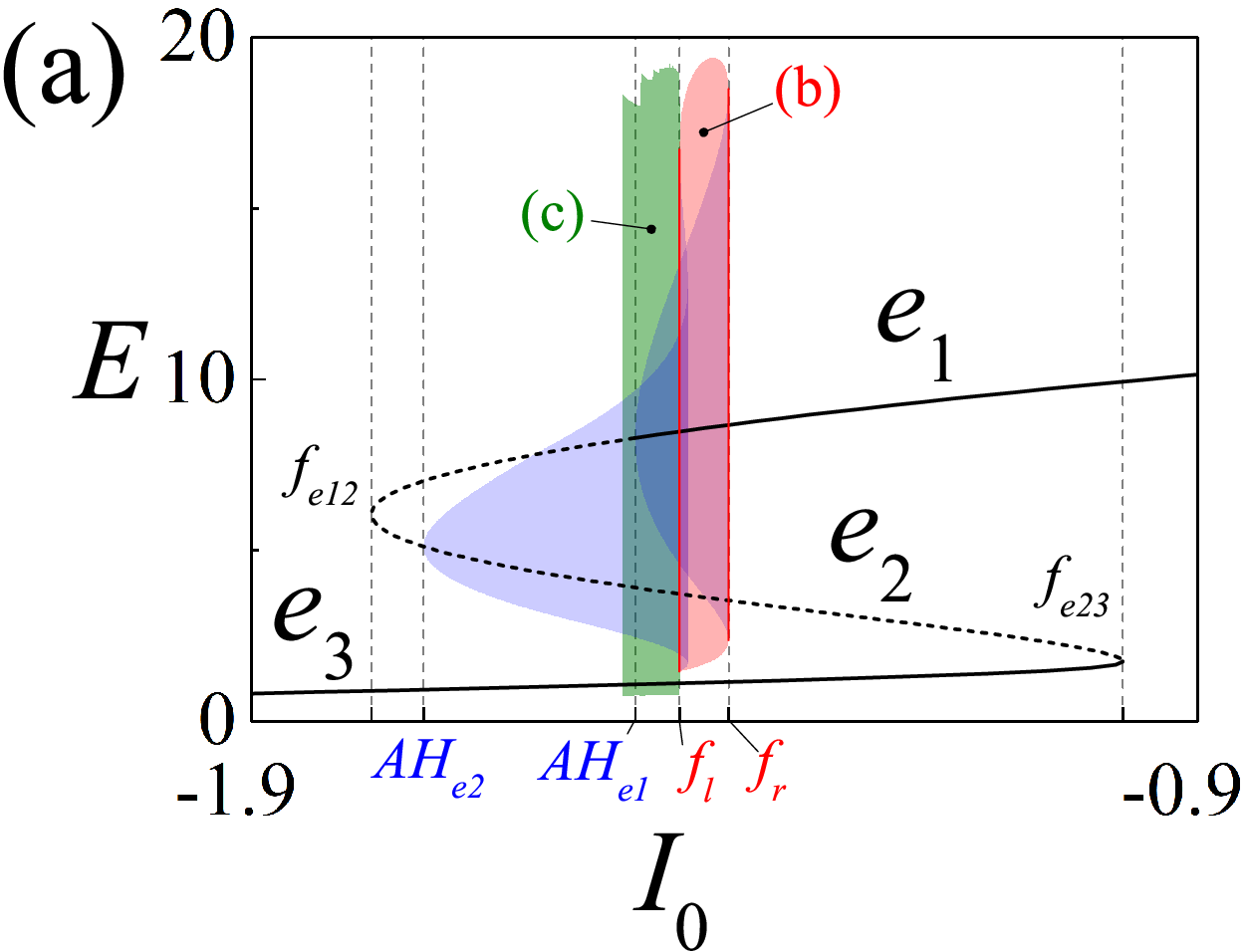}\quad\quad
    \includegraphics[width=0.35\textwidth]{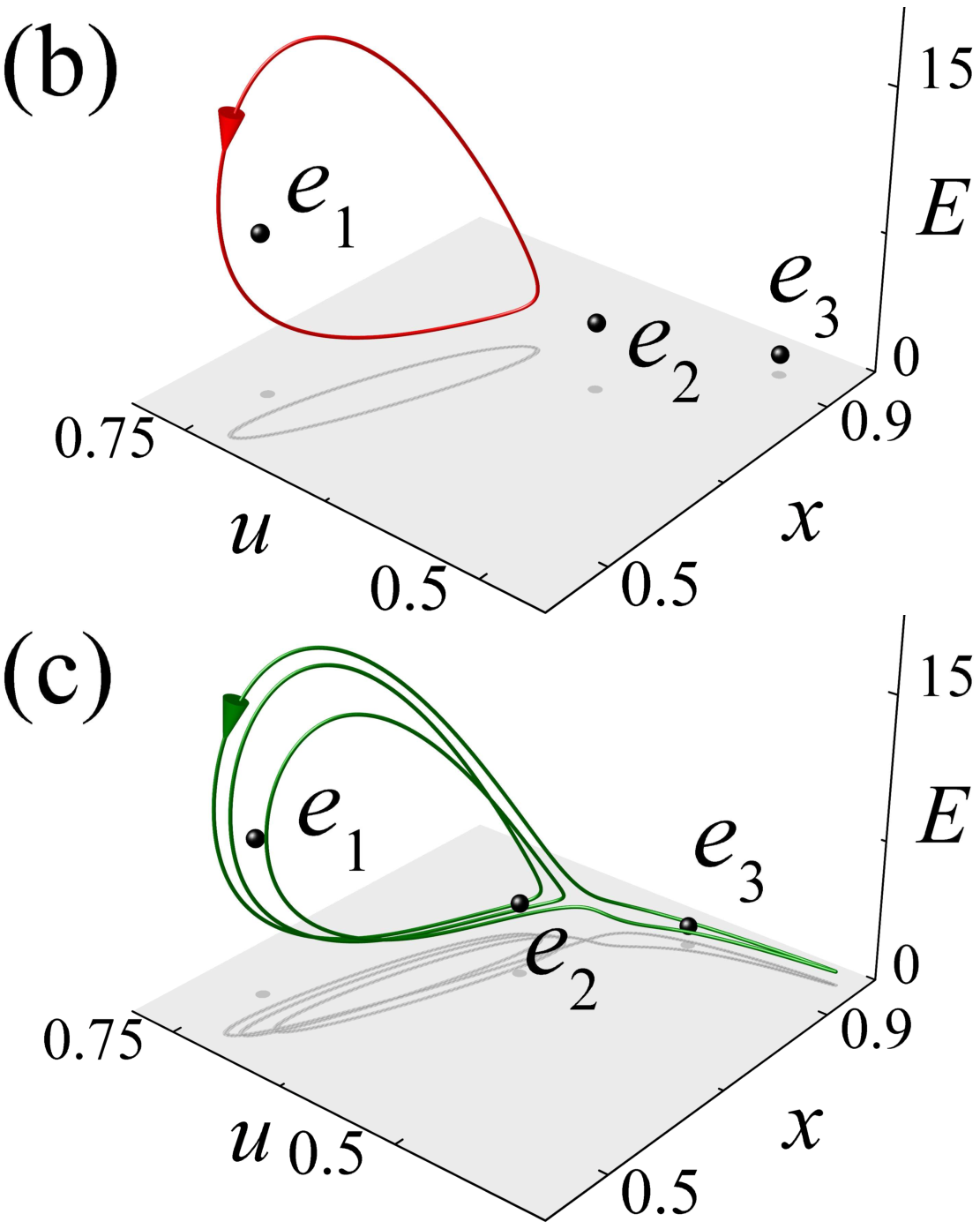}
    \caption{Bifurcation transitions explaining the change in the population activity in the system~\eqref{sys:model}. (a) Extended one-dimensional bifurcation diagram. The black curve marks the coordinate $E$ of the equilibrium states $e_1$, $e_2$ and $e_3$. Solid and dashed areas mean their stability and instability, respectively. Vertical dashed lines mark the main bifurcations: $f_{e23}$ ($f_{e12}$) is a saddle-node bifurcation of equilibrium states $e_2$ and $e_3$ ($e_1$ and $e_2$ respectively), $AH_{e1}$ and $AH_{e2}$ are the Andronov-Hopf bifurcations of $e_1$ and $e_2$, at which saddle cycles are born (the blue bell-shaped regions are their projections onto the $E$ axis), $f_l$ and $f_r$ are saddle-node bifurcations of limit cycles. The red area is the projection onto the $E$ axis of a stable limit cycle (red curve in panel (b)) corresponding to oscillatory activity. The green area is a projection of multi-pass stable cycles (green curve in panel (c)) and irregular motions corresponding to bursting activity. Panels (b) and (c) show three-dimensional projections of stable limit cycles corresponding to oscillatory ($I_0 = -1.42$) and bursting ($I_0 = -1.48$) activity, respectively. The transition from oscillatory activity to bursting activity occurs through the saddle-node bifurcation $f_l$, where the one-loop stable limit cycle (red curve in panel (b)) merges with the saddle limit cycle. 
    }
\label{fig:bif_diag}
\end{figure*}

The system~\eqref{sys:model}, depending on the parameter $I_0$, has one ($e_1$ or $e_3$) or three equilibrium states $e_{1, 2, 3}$, which appear and disappear as a result of saddle-node bifurcations.
Fig.~\ref{fig:bif_diag}(a) shows an extended one-parameter bifurcation diagram. It was built numerically and demonstrates bifurcations that take place in the case of changing control parameter $I_0$. 
The black S-shaped curve in Fig.~\ref{fig:bif_diag}(a) marks the change in the $E$ coordinate and the stability type of the equilibrium states as the parameter $I_0$ changes.
The solid and dashed parts of the curve correspond to the stability and instability of $e_{1,2,3}$, respectively.
Namely, $e_3$ is always stable, $e_2$ is always unstable, and $e_1$ changes the type of its stability under the Andronov-Hopf bifurcation $AH_{e1}$, when a saddle limit cycle merges with the stable equilibrium $e_1$.
As a result of the saddle-node bifurcations $f_{e23}$ or $f_{e12}$, the equilibrium states $e_2$ and $e_3$ (or $e_1$ and $e_2$, respectively) merge and disappear, leaving the equilibrium state $e_1$ (or $e_3$ respectively) globally stable.

Spiking and bursting dynamics of the system~\eqref{sys:model} correspond to stable periodic orbits (and in some cases, chaotic attractors, which are not considered in this article).
The one-loop stable limit cycle corresponding to regular spiking activity is born together with the saddle limit cycle as a result of saddle-node bifurcations $f_l: I_0 \approx -1.447$ and $f_r: I_0 \approx -1.396$ (see. Fig.~\ref{fig:bif_diag}(a)).
A three-dimensional projection of a stable limit cycle that exists in the system~\eqref{sys:model} at $I_0 = -1.42$ is shown in Fig.~\ref{fig:bif_diag}(b).

The transition from the spiking to the bursting population activity occurs as a result of the saddle-node bifurcation $f_l: I_0 \approx -1.447$, at which the stable limit cycle (the red closed trajectory in Fig.~\ref{fig:bif_diag}(b)) merges with the saddle limit cycle and disappears.
In this case, a stable multi-pass limit cycle of a large period is born, which corresponds to a bursting activity with a burst of long duration.
\footnote{Such transitions occur during the ``blue sky catastrophe'' bifurcation \cite{palis1975fifty,turaev1995blue,kuptsov2017family} and have been considered, incl. in the context of applications in neurodynamics \cite{Shilnikov2005}.}

With the decrease in parameter $I_0$ the size of the burst also decreases (see temporal patterns in Fig.~\ref{fig:realizations}).
In Fig.~\ref{fig:bif_diag}(c) the green closed trajectory shows the three-dimensional projection of the multi-pass limit cycle corresponding to the bursting activity at $I_0 = -1.48$.
For $I_0 \approx -1.509$ the multi-pass stable limit cycle vanishes, and the equilibrium state $e_3$ becomes the only attracting set.

Thus, the system~\eqref{sys:model} for the considered parameter values in the interval $-1.509 < I_0 < -1.396$ can have three attracting sets simultaneously, which means that the dynamics of the system depends on the initial conditions. From the biophysical point of view, described multistability at the network level could ensure the formation of persistent states in the prefrontal cortex \cite{Mongillo2008}.

\section{Conclusion}


In this study we have proposed a new mean-field model of neuron-glial interactions, which reproduces bursting activity. The novelty of our research is the extension of a classical Tsodyks-Markram model that allows to take into account the effects of glial cells to understand the mechanisms for the generation of the bursting activity in neuronal populations. 

Previously the Tsodyks-Markram model with short-term synaptic plasticity has already been successfully applied to various phenomena such as working memory \cite{Wang2006} and working memory processing capacity \cite{Mi2017}. The dynamic properties of the Tsodyks-Markram model with short-term synaptic plasticity were studied in \cite{Cortes2013}. Our model allows taking into account the important features of the glial mechanism of modulation of the probability of neurotransmitter release, leading to bursting activity. To the best of our knowledge, these peculiarities have not been considered earlier.


The developed model allows to reproduce a rich variety of temporal patterns: from trivial ones, such as quiescence and tonic spiking, to regular and irregular bursting activity. 
Mathematical images of these types of activity in the phase space of the proposed system were described. In particular, it was shown that regular bursting activity is connected to the appearance of multi-loop periodic orbits in the phase space of the system under study. We have used bifurcation theory to obtain the mathematical description of transitions between the main types of population activity, caused by variations in the control parameter $I_0$ that characterize inhibitory input. Irregular bursting activity in the system~\eqref{sys:model} is generated by a chaotic attractor. This issue is beyond the scope of this work and will be considered separately. 

The appearance of multistability and bursting activity in the model is independent of the complexity of the local dynamics of neurons and glial cells. These types of dynamics are determined by the existence of the feedback loop between presynaptic terminal and glial cells. The demonstrated effects of bursting dynamics and neuron-glial interactions are robust because they do not imply specific characteristics in the neuron–glial interaction, a particular architecture of the neural network or dynamics of individual neurons. 
  
Summarizing, the proposed phenomenological mean-field model can be used to reproduce different patterns of population spiking and bursting activity in a wide range of studies of dynamic memory and information processing. One possible application of such studies is development of new efficient treatment of neurological diseases related to neuron-glial interactions. Another area, where these results can be helpful, concerns creation of efficient living chip with useful functions, which requires better insights into rhythmogenesis in neural networks and the functioning of the brain.

{\bf Acknowledgement.}

The study was supported by a grant from the Russian Science Foundation 22-12-00348.

\section*{Declarations}
The datasets generated and/or analysed during the current study are available from the corresponding author on reasonable request.

\bibliographystyle{epj.bst}
\bibliography{biblio}

\end{document}